\documentclass[12pt,aps,prb]{revtex4} 
\usepackage{graphicx}
\usepackage{amssymb}
\usepackage{amscd}
\usepackage{mathrsfs}
\usepackage{longtable,lscape}
\usepackage{amsthm}
\usepackage{amsfonts}
\usepackage{amsmath}
\usepackage{bbm}
\usepackage{float}
\usepackage{url} 
\usepackage{graphicx}

\begin{document}

\title{On Aerts' overlooked solution to the EPR paradox}
\author{Massimiliano Sassoli de Bianchi}
\affiliation{Center Leo Apostel for Interdisciplinary Studies, Brussels Free University, Brussels, Belgium}
\email{msassoli@vub.ac.be} 

\date{\today}

\begin{abstract}
\noindent
The Einstein-Podolsky-Rosen (EPR) paradox was enunciated in 1935 and since then it has made a lot of ink flow. Being  a subtle result, it has also been largely misunderstood. Indeed, if questioned about its solution, many physicists will still affirm today that the paradox has been solved by the Bell-test experimental results, which have shown that entangled states are real. However, this remains a wrong view, as the validity of the EPR \emph{ex-absurdum} reasoning is independent from the Bell-test experiments, and the possible structural shortcomings it evidenced cannot be eliminated. These were correctly identified by the Belgian physicist Diederik Aerts, in the eighties of last century, and are about the inability of the quantum formalism to describe separate physical systems. The purpose of the present article is to bring Aerts' overlooked result to the attention again of the physics' community, explaining its content and implications.
\end{abstract}

\maketitle

\section{Introduction}

In 1935, Albert Einstein and his two collaborators, Boris Podolsky and Nathan Rosen (abbreviated as EPR), devised a very subtle thought experiment to highlight possible inadequacies of the quantum mechanical formalism in the description of the physical reality, today known as the \emph{EPR paradox}.\cite{EPR} The reason for the ``paradox'' qualifier is that the predictions of quantum theory, regarding the outcome of their proposed experiment, differed from those obtained by means of a reasoning using a very general reality criterion.

Despite the fact that the EPR objection to quantum mechanics has been the subject of countless discussions in the literature (see for instance Refs.~\onlinecite{Hooker1970,Reisler1971,Erlichson1972,Paul1985,Griffiths1987,Blaylock2010}), many physicists still believe today that the EPR paradox has been solved by the celebrated coincidence experiments on pair of entangled photons in singlet states, realized by Alain Aspect and his group in 1982,\cite{Aspect1982} which were later reproduced under always better controlled experimental situations,\cite{Aspect1999} closing one by one all potential experimental loopholes.\cite{Hensen2015} More precisely, the belief is that these experiments would have invalidated EPR's reasoning by confirming the exactness of the quantum mechanical predictions.

This conclusion, however, is the fruit of a misconception regarding the true nature of the EPR paradox, which was not solved by experiments like those conducted by Aspect et al., but by a constructive proof presented almost forty years ago by Diederik Aerts, in his doctoral dissertation.\cite{Aerts1981,Aerts1982Found,Aerts1983AVCP,Aerts1984a,Aerts1984b,Aerts1984c} Contrary to what is generally believed, Aerts' solution says that the quantum mechanical description of reality is indeed incomplete, because, as we are going to explain, it cannot describe separate physical systems. Aerts' result remains to date largely unknown, and the main purpose of the present article is to bring it back to the attention of the scientific community. I will do so by trying to explain it in the simplest possible terms, also indicating its consequences for our understanding of classical and quantum theories. 

\section{Correlations}

We start by observing that quantum entanglement, which was firstly discussed by EPR\cite{EPR} and Schr{\" o}dinger,\cite{s1935a,s1935b} is incompatible with a classical spatial representation of the physical reality. Indeed, in this representation a spatial distance also expresses a condition of \emph{experimental separation} between two physical entities, in the sense that the greater the spatial distance $\Delta x$ between two entities $A$ and $B$, and the better $A$ and $B$ will be experimentally separated. To be experimentally separated means that when we test a property on entity $A$, the outcome of the test will not depend on other tests we may want to perform, simultaneously or in different moments, on entity $B$, and vice versa.\footnote{More precisely, it will not depend on them in an ontological sense, rather, possibly, in a dynamical sense, for instance because both entities may interact by means of a force field, such as the gravitational or electromagnetic fields. In other words, quoting from Ref.~\onlinecite{Aerts1984a}: ``In general there is an interaction between separate systems and by means of this interaction the dynamical change of the state of one system is influenced by the dynamical change of the state of the other system. In classical mechanics for example almost all two body problems are problems of separate bodies (e.g. the Kepler problem). Two systems are non-separate if an experiment on one system changes the state of the other system. For two classical bodies this is for example the case when they are connected by a rigid rod.''}

For two classical entities this will be the case if $\Delta x$ and the time interval $\Delta t$ between the different tests is such that no signal can propagate in time between the two entities to possibly influence the outcomes of the respective tests, which will be the case if ${\Delta x \over \Delta t} > c$, with $c$ the speed of light in vacuum. Of course, in the limit $\Delta t\to 0$, where the two tests are performed in a perfectly simultaneous way, any finite distance $\Delta x$ will be sufficient to guarantee that we are in a non-signaling condition, i.e., that we are in a situation of experimental separation. In other words, in classical physics the notions of \emph{spatial separation} and \emph{experimental separation} were considered to be intimately connected, in the sense that the former was considered to generally imply the latter. 

\begin{figure}
\centering
\includegraphics[width=8cm]{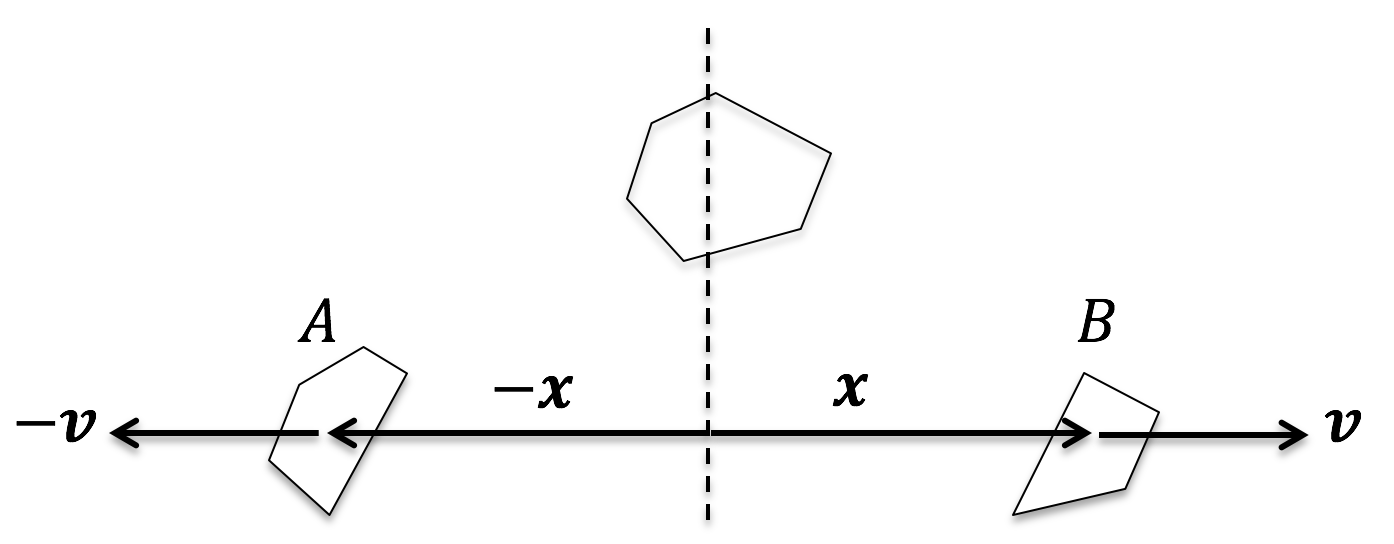}
\caption{A rock initially at rest explodes into two fragments which are here assumed to be of equal masses, flying apart in space with opposite velocities.}
\label{Figure1}
\end{figure}

Consider now that the two entities $A$ and $B$ are two bodies moving in space in opposite directions and assume that two experimenters decide to jointly measure their positions and velocities. Since the two entities are spatially separate, and therefore perfectly disconnected, no correlations between the outcomes of their measurements will in general be observed. However, if the two objects were connected in the past, the physical process that caused their disconnection may have created correlations that subsequently can be observed. As a paradigmatic example, consider a rock initially at rest, say at the origin of a laboratory's system of coordinates, and assume that at some moment it explodes into two fragments $A$ and $B$, having exactly equal masses (see Fig.~\ref{Figure1}). The positions and velocities of these two flying apart fragments of rock will then be perfectly correlated, due to the conservation of momentum: if at a given instant the position and velocity of (the center of mass of) fragment $A$ are ${\bf x}$ and ${\bf v}$, respectively, then the position and velocity at that same instant of fragment $B$ will be $-{\bf x}$ and $-{\bf v}$. This situation of perfect correlation is clearly the consequence of how the two fragments were created in the past, out of a single whole entity, and is not the result of a connection that is maintained between the two fragments while moving apart in space.

It is in fact important to distinguish between the correlations that can only be \emph{discovered}, between the two components of a bipartite system, and which are due to previous processes of connections-disconnection, from the correlations that are literally \emph{created} by the very process of their observation, i.e., which are created out of an actual connection between the two components of the bipartite system, when these two parts are subjected to a measurement. This fundamental distinction was made in the nineties by Aerts, who specifically named the correlations that are only discovered in a measurement \emph{correlations of the first kind}, and those that are instead created in a measurement \emph{correlations of the second kind}.\cite{Aerts1990}

The key role played by Bell's inequalities\cite{Bell1964,Bell1971} in identifying the presence of entanglement in composite physical systems can then be identified in their ability to demarcate between correlations of the first kind and correlations of the second kind, as only the latter can violate them. In that respect, it is important to note that the violation of Bell's inequalities is not a specificity of micro-physical systems: also classical macroscopic systems can violate them, as what is truly important for the violation is to have correlations that can be created during the very process of measurement, which will be generally the case when the two entities forming the bipartite system are connected in some way, for instance because they are in direct contact, or because of the presence of a third connecting element. So, to give examples, two vessels of water connected through a tube,\cite{Aerts1984c,Aerts1982} or two dice connected through a rigid rod,\cite{Sassoli2013,Sassoli2014} can easily violate Bell's inequalities in specifically designed coincidence experiments.\footnote{To give another remarkable example, abstract conceptual entities, which are connected through meaning, are also able to violate Bell's inequalities similarly to quantum micro-entities, in specific psychological measurements.\cite{Aertsetal 2017-I,Aertsetal 2017-II}}

But then, if classical entities can also produce quantum-like correlations of the second kind, violating Bell's inequalities, why Einstein famously called the quantum correlations ``spooky actions at a distance''? The answer is simple: a tube connecting two vessels of water, or a rod connecting two dice, are elements of reality that can be easily described in our three-dimensional Euclidean theater, so there is no mystery in their functioning, whereas what keeps two micro-physical entities connected in a genuine quantum entangled state apparently cannot. In other words, the ``spookiness'' of the quantum correlations comes from the fact that: (1) they are not correlations of the first kind and (2) the connectedness out of which the correlations are created is a \emph{non-spatial} element of our physical reality.\footnote{Interestingly, this ``quantum connectedness element,'' characterizing the potential correlations that can be actualized in a coincidence measurements with entangled entities, can be explicitly represented in the generalized quantum formalism called the Extended Bloch Representation (EBR) of quantum mechanics.\cite{AertsSassoli2016}}

\section{The paradox}

Let me now explain the EPR reasoning in their celebrated article.\cite{EPR} First of all, they introduced the important notion of \emph{element of reality}, corresponding to the following definition:\cite{EPR} ``\emph{If without in any way disturbing the state of a physical entity the outcome of a certain observable can be predicted with certainty, there exists an element of reality corresponding to this outcome and this observable}.'' Contrary to what is often stated, the important part in this definition is not the ``without in any way disturbing'' one, but the ``can be predicted with certainty'' one. Indeed, as observed by Aerts:\cite{Aerts1983AVCP} ``it is possible for an entity to have an element of reality corresponding to a physical quantity even when this physical quantity cannot be measured without disturbing the entity.''

Consider for instance the burnability property of a wooden cube.\cite{Aerts1982Found} We know it is an actual property, as we can \emph{predict with certainty} that if we put the cube on fire it will burn with certainty (i.e., with probability equal to $1$), but of course we cannot test the burnability property without deeply disturb the wooden entity. In their paper, EPR did not mention explicitly this subtle point, which however was later on integrated by Constantin Piron in his fundamental definition of an \emph{actual property}\cite{Piron1, Piron2} and used in his construction of an axiomatic operational-realistic approach to the foundations of quantum mechanics. According to this definition, a property is actual if and only if, should one decide to perform the experimental test that operationally defines it, the expected result would be certain in advance. If this is the case, the entity in question is said to have the property (i.e., to possess it in actual terms) even before the test is done, and in fact even before one has chosen to do it (independently on the fact that the test might be invasive or not). And this is the reason why one is allowed to say that the property is an element of reality, existing independently from our observation. 

So, even if not fully expressed at the time, the EPR reasoning contained the deep insight that the (actual) properties of physical systems are ``states of predictions.'' EPR then considered the situation of two quantum entities $A$ and $B$ that, after interacting, subsequently flew apart in space, becoming in this way spatially separate and, according to EPR's prejudice, also experimentally separate. The additional step taken by EPR in their 1935 paper is to consider the quantum mechanical formalization of this situation, in accordance with the notion of entanglement, from which they observe that the positions and velocities of the two quantum entities are strongly correlated. 

More precisely, the EPR reasoning goes as follows. They consider the possibility of measuring the position of one of the quantum entities, say entity $B$ that is flying to the right. Assuming that such measurement has been carried out, and that the position of entity $B$ has been observed to be ${\bf x}$, then, according to the quantum description, the experimenter is in a position to predict that if a position measurement would be performed on entity $A$, the outcome $-{\bf x}$ would be obtained with certainty (considering a system of coordinates such that the place where the two entities interacted before flying apart corresponds to its origin). The subtle point here is that since $A$ and $B$ are separated by an arbitrarily large spatial distance, and that the assumption is that a spatial separation also implies an experimental separation, the previous measurement on $B$ could not affect in whatsoever way the state of $A$. Hence, the prediction that the position of $A$ is $-{\bf x}$ establishes the actuality of the property, and of course the same reasoning holds in case it is the velocity (or momentum) that is measured on $B$, as also in this case, if the outcome of the measurement was, say, ${\bf v}$, then the outcome $-{\bf v}$ could have been predicted with certainty for entity $A$. 

Is the above sufficient to conclude that entity $A$ has both a well-defined position and velocity? To clarify the situation, let me come back to Aerts' example of the wooden cube, which as we observed has the property of being burnable. We also know that it has other properties, like the property of floating on water. How do we know that? Again, because if we would perform the test of immersing the cube in water, the ``floating on water'' outcome would be obtained with certainty. But then, we can ask the following question: Does the wooden cube jointly possess the properties of burnability and floatability? Our common sense tells us that this has to be the case, but how do we test this \emph{meet property} obtained by the combination of the burnability and floatability properties? Because a wet cube does not burn, and a burned cube does not float, so we cannot conjunctly or sequentially test these two properties. 

In fact, we don't have to, because, as noted by Piron,\cite{Piron1} the test for a meet property is a so-called \emph{product test}, consisting in performing only one of the two tests, but chosen in a random (unpredictable) way. Indeed, the only way we can then predict the positive outcome of a procedure consisting in randomly selecting one of the two tests, then executing it, is to be able to predict the positive outcome of both tests, which precisely corresponds to the situation where both properties are simultaneously actual. 

Having clarified that even when the experimental tests of two properties are mutually incompatible this does not imply that they cannot be jointly tested by means of a product test, and therefore the properties be simultaneously actual, we can now observe that the EPR reasoning precisely describes a situation where the outcome of a product test (or product measurement) for the position and velocity of entity $A$ can be predicted with certainty. Indeed, in case it is the position measurement that is randomly selected, the experimenter can perform that same measurement on entity $B$ and then predict with certainty the outcome of the position measurement on entity $A$, without the need to perform it. And in case it is the velocity (or momentum) measurement that is randomly selected, the experimenter can perform the velocity measurement on entity $B$ and again predict with certainty the outcome of the same measurement on entity $A$, again without the need to perform it. In other words, we are exactly in a situation where the outcome of a product measurement of position and velocity observables can be predicted with certainty, hence, we are allowed to conclude, with EPR, that both position and velocity have simultaneous well-defined values for entity $A$. This is of course in flagrant contradiction with Heisenberg's uncertainty relations, hence the paradox and EPR's conclusion that quantum mechanics is an incomplete theory, as unable to represent all possible elements of reality associated with a physical entity. 

Bohr's reaction to the EPR argument, that same year, was quite obscure.\cite{Bohr1935} Basically, the Danish physicist affirmed that one ``is not allowed in quantum mechanics to make the type of reasoning proposed by EPR, and more specifically, the notion of element of reality does not make sense for quantum mechanical entities.'' With the exception of Schr{\" o}dinger, Bohr's authority (and the influence of the Copenhagen interpretation) resulted in most leading quantum physicists simply accepting that there was not really a serious problem involved in the EPR reasoning and resulting paradox. Many years later though, perhaps because also of the influence of David Bohm, who certainly took the EPR argument seriously (inventing the entangled spin example as a more transparent description of the EPR situation), a small group of physicists, among whom was John Bell, believed that EPR highlighted a fundamental problem in quantum mechanics related to its possible incompleteness. However, different from what EPR, Bohm, Bell and others believed, the incompleteness in question was not an issue of ``providing additional variables'' to make it complete, or more complete, but a question of a shortcoming related to the impossibility for the quantum formalism to describe experimentally separate entities, as subsequently shown by Aerts.\cite{Aerts1981,Aerts1982Found,Aerts1983AVCP,Aerts1984a,Aerts1984b,Aerts1984c}

\section{The solution}

To explain Aerts' solution, it is important to emphasize that EPR's reasoning is an \emph{ex absurdum} one, that is, a reasoning which starts from certain premises and reaches a contradiction. What EPR have shown is that if their premises are assumed to be correct, then quantum theory has to be considered incomplete, as unable to describe all elements of reality of a physical system. Those who have taken seriously this conclusion thus tried to find remedies, for instance by supplementing the theory with additional variables for the quantum states, to allow position and velocity to have simultaneous definite values and escape the limitation of Heisenberg's uncertainty relations. This hidden variables program, however, subsequently met the obstacle of so-called \emph{no-go theorems}, drastically limiting the class of admissible hidden-variable theories.\cite{Neumann1932,Bell1966,Gleason1957,Jauch1963,Kochen1967,Gudder1970} 

The premise that was part of the EPR reasoning, as we explained, is that for two quantum entities that have interacted and flown apart, it was natural to expect that their spatial separation was equivalent to an experimental separation. In addition to that, EPR applied the quantum formalism to describe the situation, which means they implicitly also assumed that quantum mechanics is able to describe a system formed by \emph{separate physical entities}. But since this produced a contradiction, one is forced to conclude that the assumption is incorrect, that is, that quantum mechanics is unable to describe separate entities. 

Now, one may object that this is a too strong conclusion, in the sense that the only mistake committed by EPR was to expect that spatial separation would also necessarily imply disconnection. This expectation, as we know today, has been overruled by numerous experiments, showing that by making sufficient efforts and taking all necessary precautions, experimental situations can indeed be created where microscopic entities, after having interacted, can remain interconnected, even when arbitrarily large spatial distances separate them. The mistake of EPR was therefore to think about a situation where there is no experimental separation between two entities, as a situation of actual experimental separation. 

So, apparently problem solved: EPR-like experiments, like those performed by the group of Alain Aspect, have precisely shown that in the situation considered by EPR quantum mechanics does actually provide the correct description of two quantum entities flying apart, since Bell's inequalities are violated, in accordance with the quantum predictions. Thus, one would be tempted to conclude that EPR's reasoning is not valid. Well, yes and no. Yes, because at their time the possibility of producing these non-local/non-spatial states was a truly remarkable and totally unexpected possibility, based on classical prejudices, so the EPR \emph{ex absurdum} reasoning was indeed applied to a wrong experimental situation, if such situation is considered to be correctly described by an entangled state. No, because the possibility of producing and preserving entangled states has very little to do with EPR's reasoning per se. Indeed, one can in principle also assume that experiments could be performed where instead of making efforts to preserve the quantum connectedness of the two flying apart entities, an effort is made instead to obtain the opposite situation of two flying apart entities eventually becoming perfectly disconnected, i.e., separated. 

Experiments of this kind have never been worked out consciously, but these would indeed correspond to situations leading to the EPR paradox. In other words, the incompleteness of quantum mechanics is not revealed in the physical situation of quantum entities flying apart and remaining non-separate, as these are the situations which are perfectly well described by the quantum formalism (as the violation of Bell's inequalities proves), and there is no contradiction/paradox in this case, but by the experimental situations that can produce a disconnection, and which in the setting of EPR-like experiments would be interpreted as ``badly performed experiments.'' These are precisely the situations that quantum mechanics would be unable to describe, certainly not by means of entangled states, as if we assume it can, then we reach a contradiction.

Having clarified that the logical reasoning of EPR is not directly affected by the experimental discovery of entangled states, the question thus remains about the completeness of the quantum formalism, in relation to its ability to describe separate physical entities. It is here that Aerts' work join the game. Indeed, among the topics of his doctoral research there was that of elaborating a mathematical framework for the general description of separate quantum entities. Aerts approached the issue using Piron's axiomatic approach to quantum mechanics, a very general formalism which was precisely:\cite{Piron3} ``obtained by taking seriously the realistic point of view of Einstein and describing a physical system in terms of `elements of reality'.'' This allowed him to view the EPR work from a completely new angle. Indeed, while describing the situation of bipartite systems formed by separate quantum entities, he was able to prove, this time in a perfectly constructive way, that quantum mechanics is structurally unable to describe these situations.

\section{Aerts' proof}

EPR were thus right about the incompleteness of quantum mechanics, but not for the reason they believed: quantum mechanics is incomplete because unable to describe separate physical systems. Of course, depending on the viewpoint adopted, this can be seen as a weak or strong trait of the theory. If separate systems exist in nature, then it is a weak trait, if they don't, then it is a strong trait. We will come back on that in the conclusive section, but let us now sketch the content of Aerts' constructive proof, which is actually quite simple. 

Note that despite the simplicity of the proof, it usually comes as a surprise that quantum mechanics would have this sort of shortcoming. Indeed, the first reaction I usually get, when discussing Aerts' result with colleagues, is that this cannot be true, as separate systems are perfectly well described in quantum mechanics by so-called product states, that is, states of the tensor product form $\psi\otimes\phi$, where $\psi\in {\cal H}_A$ and $\phi\in {\cal H}_B$, with ${\cal H}_A$ the Hilbert (state) space of entity $A$ and ${\cal H}_B$ that of entity $B$, the Hilbert space ${\cal H}$ of the bipartite system formed by $A$ and $B$ being then isomorphic to ${\cal H}_A\otimes{\cal H}_B$. This is correct, and in fact the shortcoming of quantum theory in describing separate systems cannot be detected at the level of the states, as in a sense there is an overabundance of them, but at the level of the properties, which in the quantum formalism are described by orthogonal projection operators. In fact, it is precisely this overabundance of states that produces a deficiency of properties, in the sense that certain properties of a bipartite system formed by separate components cannot be represented by orthogonal projection operators.

Technically speaking, the only difficulty of Aerts' proof is that one needs to work it out in all generality, independently of specific representations, like the tensorial one, so that one can be certain that its conclusion is inescapable.\cite{Aerts1981,Aerts1982Found,Aerts1983AVCP,Aerts1984a,Aerts1984c} Without entering into all details, the demonstration goes as follows. First, one has to define what it means for two entities $A$ and $B$ to be experimentally separate. As we mentioned already, this means that measurements individually performed on them do not influence each other. In other words, separate entities are such that their measurements are \emph{separate measurements}. More precisely, two measurements ${\cal M}_A$ and ${\cal M}_B$ are separate if they can be performed together without influencing each others. This means that, from them, one can define a combined measurement ${\cal M}_{AB}$ such that: (1) the execution of ${\cal M}_{AB}$ on the bipartite entity formed by $A$ and $B$ corresponds to the execution of ${\cal M}_A$ on $A$ and of ${\cal M}_B$ on $B$, and (2) the outcomes of ${\cal M}_{AB}$ are given by all possible couples of outcomes obtained from ${\cal M}_A$ and ${\cal M}_B$. 

What Aerts then shows is that there is no self-adjoint operator $O_{AB}$ that can represent such measurement ${\cal M}_{AB}$. To do so, he considers two arbitrary projections $P_A^I$ and $P_B^J$, in the spectral decomposition of the self-adjoint operators $O_A$ and $O_B$ associated with measurements ${\cal M}_A$ and ${\cal M}_B$, respectively. Here $I$ and $J$ are subsets of the outcome sets $E$ and $F$ of the two measurements, respectively. He also defines the spectral projection $P_{AB}^{I\times J}$ of $O_{AB}$, where $I\times J$ is the subset of the outcome sets of ${\cal M}_{AB}$ formed by all couples $(x,y)$ of elements $x\in I$ and $y\in J$. Then he shows (we do not go into the details of this here), as one would expect, that $[P_A^I, P_B^J]=0$, so that also $[O_A, O_B]=0$, and that $P_{AB}^{I\times J}=P_A^IP_B^J$.

The next step is to consider a state $\psi\in {\cal H}$ which can be written as a superposition $\psi = {1\over\sqrt{2}}(\phi + \chi)$, where $\phi$ belongs to the subspace $P_A^I(\mathbb{I}-P_B^J){\cal H}$ and $\chi$ to the subspace $(\mathbb{I}-P_A^I)P_B^J{\cal H}$, orthogonal to the latter. It follows that:
\begin{eqnarray}
&&P_A^I\psi={1\over\sqrt{2}}\phi, \quad (\mathbb{I}-P_A^I)\psi={1\over\sqrt{2}}\chi, \nonumber\\
&&P_B^J\psi={1\over\sqrt{2}}\chi, \quad (\mathbb{I}-P_B^J)\psi={1\over\sqrt{2}}\phi. 
\end{eqnarray}
This means that when the bipartite system is in state $\psi$, there is at least two possible outcomes $x_1\in I$ and $x_2\in E- I$, for measurement ${\cal M}_A$, and at least two possible outcomes $y_1\in J$ and $y_2\in F- J$, for measurement ${\cal M}_B$. This means that the four outcomes $(x_1,y_1)\in I\times J$, $(x_1,y_2)\in I\times (F- J)$, $(x_2,y_1)\in (E- I)\times J$ and $(x_2,y_2)\in (E- I)(F- J)$ should be all possible outcomes of measurement ${\cal M}_{AB}$, if ${\cal M}_A$ and ${\cal M}_B$ are assumed to be separate measurements. But although we have: 
\begin{eqnarray}
&&P_{AB}^{I\times (F-J)}\psi=P_A^I(\mathbb{I}-P_B^J)\psi=\phi, \nonumber\\
&&P_{AB}^{(E-I)\times J}\psi=(\mathbb{I}-P_A^I)P_B^J\psi=\chi,
\end{eqnarray}
so that $(x_1,y_2)$ and $(x_2,y_1)$ are possible outcomes of ${\cal M}_{AB}$, we also have that: 
\begin{eqnarray}
&&P_{AB}^{I\times J}\psi=P_A^IP_B^J\psi=0, \nonumber\\
&& P_{AB}^{(E-I)\times (F-J)}\psi=(\mathbb{I}-P_A^I)(\mathbb{I}-P_B^J)\psi=0.
\end{eqnarray}
Hence, $(x_1,y_1)$ and $(x_2,y_2)$ are not possible outcomes of ${\cal M}_{AB}$, which means that ${\cal M}_A$ and ${\cal M}_B$ are not separate measurements. 

In other words, because of the superposition principle, a joint measurement ${\cal M}_{AB}$ formed by two separate measurements ${\cal M}_A$ and ${\cal M}_B$ cannot be consistently described in quantum mechanics, which means that quantum mechanics, for structural reasons related to its vector space structure, cannot handle separate measurements. 

Note that when one introduces the more specific tensorial representation ${\cal H}={\cal H}_A\otimes {\cal H}_B$, the request for the self-adjoint operators associated with measurements ${\cal M}_A$ and ${\cal M}_B$ to commute is automatically implemented by writing them in the tensorial form $O_A \otimes \mathbb{I}_B$ and $\mathbb{I}_A\otimes O_B$, respectively, so that we also have in this case $P_{AB}^{I\times J}=P_A^I\otimes P_B^J$, and the superposition state $\psi$ can for instance be written as an entangled state $\psi {1\over\sqrt{2}}(\phi_A\otimes \phi_B + \chi_A\otimes \chi_B)$, with $\phi_A\in P_A^I{\cal H}_A$, $\phi_B\in (\mathbb{I}-P_B^J){\cal H}_B$, $\chi_A\in (\mathbb{I}-P_A^I){\cal H}_A$ and $\chi_B\in P_B^J{\cal H}_B$, thus making explicit the connection of Aerts' proof with EPR-like situations.

\section{Discussion}

Having provided the gist of Aerts' demonstration, I can conclude with a few important comments. First of all, I would like to highlight once more the importance of distinguishing the logic of the EPR reasoning, leading to a paradox (contradiction), from the subsequent Bell-test experiments, the validity and interest of EPR's \emph{ex absurdum} reasoning being independent of the experimental violations of Bell's inequalities. To make this point even clearer, let me describe a different paradox, that Einstein and collaborators could have worked out at the time as an alternative reasoning to point to a possible incompleteness of quantum theory. For this, let me come back to the wooden cube and its properties of burnability and floatability. People confronted with the problem of designing an experiment able to test the joint actuality of these two properties, despite the experimental incompatibility of their individual tests, after some moments of reflection might come to the following proposal: take two additional cubes, identical to the one in question, then test the burnability on one and the floatability on the other. If both tests are successful, one can affirm that the cube under consideration jointly possess these two properties (i.e., that the meet property of ``burnability and floatability'' is actual for it). 

This is of course a possible way out to the problem of having to deal with procedures that are experimentally incompatible, so EPR could also have considered this line of reasoning to try to make their point. More precisely, they could have considered the possibility to make two identical copies of the quantum entity under investigation, measure the position on the first copy and the momentum on the second one, then present the argument that they can predict in this way, with certainty, these same values for the entity under consideration (the one that was perfectly copied), thus showing again a contradiction with Heisenberg's uncertainty principle.\footnote{The reasoning using identical copies becomes of course more convincing if expressed in relation to Bohm's version of the EPR-type situation.} Of course, since quantum measurements appear to be non-deterministic, this argument, to be valid, requires the duplication process to be ``dispersion free,'' that is, such that possible hidden variables determining the measurement outcomes are also assumed to be faithfully copied in the process. 

The reader may object that this is an invalid reasoning because of the celebrated \emph{quantum no-cloning theorem},\cite{Dieks1982,Wootters1982} establishing the impossibility of making a perfect copy of a quantum state.\footnote{Interestingly, the no-cloning theorem was proven \emph{ante litteram} by Park in 1970,\cite{Park1970} when precisely investigating the possibility of achieving a universal non-disturbing measurement scheme.\cite{Ortigoso2018}} The no-cloning theorem, however, only concerns universal copying machines, working independently of any a priori knowledge of the state to be cloned, and if we relax this condition, which we do not need for the argument, then the cloning can always in principle be worked out.\cite{Buzek1996} So, EPR could also have concluded in this case that quantum mechanics is incomplete, and once more the incompleteness cannot be associated with its inability to jointly attach position and velocity elements of reality to a micro-entity, but with its inability here of describing a perfect cloning process, when the (hypothetical) hidden variables associated with the state to be copied are unknown. 

In other words, from the above reasoning one can deduce a hidden variables variant of the no-cloning theorem: no machine can copy unknown hidden variables. Is this to be understood as an additional shortcoming of the quantum formalism? Not really, because we understand today the reason for this impossibility: hidden variables of this kind (delivering a deeper description of the reality of a physical entity) simply do not exist,\cite{Neumann1932,Bell1966,Gleason1957,Jauch1963,Kochen1967,Gudder1970} so, they cannot be copied, as of course we cannot copy what does not exist.\footnote{Note that although the no-go theorems tell us that there are no hidden variables associated with a quantum state, hidden variables can nevertheless be attached to the measurement interactions, in what was called the hidden-measurement interpretation of quantum mechanics, an approach initiated by Aerts in the eighties of last century which remains today a viable line of investigation (see Refs.~\onlinecite{AertsSassoli2016} and \onlinecite{AertsSassolideBianchi2014}, and those cited therein).} \emph{Mutatis mutandis}, Aerts' result can be understood as a \emph{quantum no-separating theorem}, establishing the impossibility of separating two physical entities, or more generally of separating the measurements (or experimental tests) associated with two physical entities. Again, we can ask: Should this be considered as a shortcoming of the quantum formalism?  

Well, maybe, as to consistently talk about a physical entity, and do physics, one must be able to consider it as a phenomenon that is separate from the rest of the universe.\cite{Aerts1982Found} Consequently, any physical entity which belongs to ``the rest of the universe'' of that physical entity, will also have to be considered to be separate from it. But this is precisely a situation that cannot be consistently described by standard quantum mechanics. The \emph{quantum measurement problem} could also be related to this limitation of the orthodox formalism, as in a measurement process the measured entity has to be initially separated from the measurement apparatus, enter into contact and interact with it, thus connect with it, then finally be separated again from it. If this connection-separation process cannot be properly described, the only way out seems that of reverting to a many-worlds picture/interpretation,~\cite{Everett,AertsSassolideBianchi2014} where separations are introduced at the level of the universes (superposition states being then described as collections of collapsed states in different universes), a move that surely would not have pleased friar Occam. 

Another difficulty one can consider, consequence of the limitations expressed by this structural impossibility of separating measurements 
and therefore entities, is in relation to  the study of (mesoscopic) structures that are in-between the quantum and classical regimes, and the quantum-classical limit. Indeed, one would need for this a more general mathematical structure for the lattice of properties than that inherited from Hilbert space and the Born rule, able to integrate both classical and quantum features. This in turn means dispensing with two of the axioms of orthodox quantum mechanics, in its lattice approach, called \emph{weak modularity} and \emph{covering law}.\cite{Piron1,Piron2,Aerts1982Found} Quoting from Ref.~\onlinecite{Aerts1999}: ``A new theory dispensing with these two axioms would allow for the description not only of structures which are quantum, classical, mixed quantum-classical, but also of intermediate structures, which are neither quantum nor classical. This is then a theory for the mesoscopic region of reality, and we can now understand why such a theory could not be built within the orthodox theories, quantum or classical.''

One can of course object that what quantum mechanics has really shown us is that all in our physical reality is deeply interconnected, that is, entangled, and that separation would be an illusion or, better, something like an effect emerging from a fundamentally interconnected non-spatial substratum, described in a correct and complete way by quantum mechanics. This is of course a possibility, although not all physicists seem to be ready to accept all the consequences of it, like the one previously mentioned of resorting to parallel universes. We live surrounded by macroscopic entities which apparently do not show quantum effects, i.e., for which separate experimental tests can be defined. If we test a property of a wooden cube, this will not influence in whatsoever way a test we may want to perform on another wooden cube. But this cannot be generally true if the Hilbertian formalism and associated superposition principle is believed to be universal. Of course, to put two wooden cubes in a state such that experiments performed on them would not anymore be separate appear to be extremely difficult to achieve, but it remains a possibility if the standard quantum formalism is considered to be fundamental. 

I personally believe that we do not know enough about our physical world to take a final stance on those difficult questions, so I think it is important to also have the possibility of studying the behavior of the different physical entities (and I stress again that the very notion of ``physical entity" requires a notion of separability) in a theoretical framework which does not attach any a priori fundamental role to the linear Hilbert space structure and associated Born rule, particularly when addressing challenging scientific problems like the one of finding a full-fledged quantum gravity theory.

\end{document}